\documentclass[aps,prd,onecolumn,preprint,groupedaddress,showpacs,nofootinbib,amssymb]{revtex4}
\usepackage[T1]{fontenc}
\usepackage{graphicx}
\usepackage[english]{babel}
\usepackage{amsmath}
\usepackage{amssymb}
\usepackage{amsfonts}

\usepackage{graphics}
\usepackage[english]{babel}
\usepackage{txfonts}
\usepackage{amsmath}
\usepackage{amsfonts}
\usepackage{amssymb}

\newcommand{\be}{\begin{equation}}
\newcommand{\ee}{\end{equation}}
\newcommand{\ba}{\begin{eqnarray}}
\newcommand{\ea}{\end{eqnarray}}
\newcommand{\bc}{\begin{center}}
\newcommand{\ec}{\end{center}}

\begin{document}

\title{Noether symmetry approach in phantom quintessence
cosmology}


\date{\today}

\author{S. Capozziello, E. Piedipalumbo, C. Rubano, and P. Scudellaro}
\affiliation{Dipartimento di Scienze Fisiche, Università di Napoli
{}`` Federico II'' and INFN Sez. di Napoli,  Complesso
Universitario di Monte S. Angelo, Via Cinthia, Edificio N', 80126
Napoli, Italy}

\begin{abstract}
In the framework of phantom quintessence cosmology, we use the
Noether Symmetry Approach to obtain general exact solutions for
the cosmological equations. This result is achieved  by  the
quintessential (phantom)  potential determined by the existence of
the symmetry itself. A comparison between the theoretical model
and observations is worked out. In particular, we use type Ia
supernovae and   large scale structure parameters determined from
the 2-degree Field Galaxy Redshift Survey (2dFGRS)and from the
Wide part of the VIMOS-VLT Deep Survey (VVDS). It turns out that
the model is compatible with the presently available observational
data. Moreover we extend the approach to include radiation. We
show that it is  compatible with data derived from recombination
and it seems that quintessence do not affect nucleosynthesis
results.
\end{abstract}


\maketitle


\section{Introduction}

Recent analysis of the three year WMAP data \cite{w,m,p} provides
no indication of any significant deviations from Gaussianity and
adiabaticity of the CMBR power spectrum and therefore suggests
that the Universe is spatially flat to within the limits of
observational accuracy. Further, the combined analysis of the
three-year WMAP data with the Supernova Legacy Survey (SNLS), in
\cite{w}, constrains the equation of state $w_{de}$, corresponding
to almost ${74\%}$ of dark energy present in the currently
accelerating Universe, to be very close to that of the
cosmological constant value. The marginalized best fit values of
the equation of state parameter gave  $-1.14 \le w_{de} \le -0.93$
at $68\%$ confidence level. Thus, it was realized that a viable
cosmological model should admit a dynamical equation of state that
might have crossed the \textit{phantom} value $w= -1$, in the
recent epoch of cosmological evolution. Phantom fluid was first
investigated in the current cosmological context by Caldwell
\cite{caldwell99}, who also suggested the name referring to the
fact that phantom (or ghost) must possess negative energy which
leads to instabilities on both classical and quantum level
\cite{instabilities,nojiri04}. Since it violates the energy
conditions, it also could put in doubt the pillars of general
relativity and cosmology such as: the positive mass theorems, the
laws of black hole thermodynamics, the cosmic censorship, and
causality  \cite{he,visser}. On the other hand,  phantom becomes a
real challenge for the theory, if its support from the supernovae
Ia-Type (SNeIa) data is really so firm. From the theoretical point
of view, a release of the assumption of an analytic equation of
state which relates energy density and pressure and does not lead
to energy conditions violation (except for the dominant one) may
also be useful \cite{barrow04}. As for the explanation of the
SNeIa data, phantom is also useful in killing the doubled positive
pressure contribution in several braneworld models \cite{braneIa}.

Phantom type of matter was also implicitly suggested in
cosmological models with a particle production \cite{john88}, in
higher-order theories of gravity models \cite{Pollock88},
Brans-Dicke models, in non-minimally coupled scalar field theories
\cite{starob00, mareknmc1}, in "mirage cosmology" of the
braneworld scenario \cite{kiritsis}, and in kinematically-driven
quintessence (k-essence) models \cite{chiba,sanyal}, for example.
Such phantom models have well-known problems but, nevertheless,
have also been widely studied as potential dark energy candidates,
and actually the interest in phantom fields has grown vastly
during the last years and various aspects of phantom models have
been investigated \cite{Frampt02,Frampt03,trodden1,
trodden2,Bastero01,Bastero02,abdalla04,Erickson,
LiHao03,singh03,nojiri031,nojiri032,nojiri033,nojiri034,simphan,onemli,diaz,saridakis1}.

One of the most interesting features of phantom models is that
they allow for a Big-Rip (BR) curvature singularity, which appears
as a result of having the infinite values of the scale factor
$a(t) \to \infty$ at a finite future. However, as it was already
mentioned, the evidence for phantom from observations is mainly
based on the assumption of the barotropic equation of state which
tightly constraints the energy density and the pressure. It is
puzzling \cite{barrow04} that for Friedmann cosmological models,
which do not admit an equation of state which links the energy
density $\varrho$ and the pressure $p$, a sudden future
singularity of pressure may appear. This is a singularity of
pressure only, with finite energy density which has an interesting
analogy with singularities which appear in some inhomogeneous
models of the Universe \cite{dabrowski95,inhosfs}.

Recently, phantom cosmologies which lead to a quadratic polynomial
in canonical Friedmann equation have been investigated
\cite{phantom1}, showing that  interesting dualities exist between
phantom and ordinary matter models which are similar to dualities
in superstring cosmologies \cite{meissner91,superjim}. These
dualities were generalized to non-flat and scalar field models
\cite{lazkoz1,lazkoz2,lazkoz3,lazkoz4,lazkoz5,lazkoz6,lazkoz7,lazkoz8,lazkoz9,lazkoz10,lazkoz11,lazkoz12,lazkoz13},
brane models \cite{calcagni}, and are also related to ekpyrotic
models \cite{ekpyrotic,triality}. Furthermore, some theoretical
studies have been devoted to shed light on phantom dark energy
within the quantum gravity framework, since, despite the lack of
such a theory at present, we can still make some attempts to probe
the nature of dark energy according to some of its basic
principles \cite{saridakis2}.

Finally, phantom cosmology can provides the opportunity to
"connect" the phantom driven (low energy $meV$ scale) dark energy
phase to the (high energy GUT scale) inflationary era. This is
possible because the energy density increases in phantom
cosmology. Concrete models in this sense have been recently
elaborated with some interesting results \cite{biswas}.

In this paper, we want to investigate if the existence of phantom
fields can be connected to Noether symmetries. Such an
issue becomes recently extremely important due to the fact that
several phenomenological models have been constructed  but, some
of them, have no self-consistent theoretical foundation. The idea
to derive the equation of state from symmetries is not new
\cite{heller} and recently has been applied to dark energy
\cite{szydlowski}. From a mathematical point of view, the general
consideration is that symmetries greatly  aid in finding exact
solutions \cite{heller,cimento}. Besides, due to the Noether
theorem, symmetries are always related to conserved quantities
which, in any case, can be considered as conserved "charges".
Specifically, the form of the self-interacting scalar-field
potential  is "selected" by the existence of a symmetry and then
the dynamics can be controlled. The equation of state, being
related to the form of scalar-field potential, is determined as
well. However, the symmetry criterion is not the only that can be
invoked to discriminate  physically consistent models but it could
be considered a very straightforward one since, as we will see
below, it allows also to achieve exact solutions.

In Sect.II, we actually show that phantom fields come out by
requiring the existence of Noether symmetry to the  Lagrangian
describing a \textit{standard} single scalar field quintessential
cosmological model: we show that it allows a phantom dark energy
field, and also provides an explicit form for the (phantom)
self-interaction potential. Sect. III studies how this gives rise
to exact and general solutions. Also extending the approach to
include radiation, we show that it is also compatible with the
post recombination observational data and that quintessence does
not influence the results of nucleosynthesis. In Sect IV, we  work
out a comparison between the theoretical solution and
observational dataset, as the publicly available data on SNeIa,
the parameters of large scale structure determined from the
2-degree Field Galaxy Redshift Survey (2dFGRS)and from the Wide
part of the VIMOS-VLT Deep Survey (VVDS). In Sect.V, we discuss
the presented results and draw conclusions.

\section{ The Noether Symmetry Approach}
The Noether Symmetry Approach has revealed a useful tool in order
to find out exact solutions, in particular in cosmology
\cite{cimento,cqgarturo,defelice,leandros}. The existence of the
Noether symmetry allows to reduce the dynamical system that, in
most of cases, results integrable. It is interesting to note that
the self-interacting potentials of the scalar field
\cite{leandros}, the couplings \cite{cimento} or the overall
theory \cite{defelice}, if related to a symmetry (i.e. a conserved
quantity) have a physical meaning. In this sense, the Noether
Symmetry Approach is also a physical criterion to select reliable
models (see \cite{defelice} for a discussion).

In the present case, let us consider a matter--dominated model in
homogeneous and isotropic cosmology with  signature $(-,+,+,+)$
for the metric, with a single scalar field, $\phi$, minimally
coupled to the gravity. It turns out that the point-like
Lagrangian
 action takes the  form
\begin{equation}
{\cal L }= 3 a \dot{a}^2 - a^3  \left(\epsilon \frac{\dot{\phi}^2}{ 2} - V(\phi)\right) + D a^{-3(\gamma -1)}\,\label{eq:lagrangian}
\end{equation}
where $a$ is the scale factor and the constant $D$ is a constant
defined in such a way that the matter density $\rho_{m}$ is
expressed as $\rho_m= D (a_o/ a)^{3\gamma}$, where $1 \leq \gamma
\leq 2$. For the moment, we will limit our analysis to $\gamma
=1$, corresponding to cosmological dust. The value of the constant
$\epsilon$ discriminates between standard and \textit{phantom}
quintessence fields: in the former case, it is $\epsilon = 1$; in
the latter, it is $\epsilon = -1$. The effective pressure and
energy density of the $\phi$-field are given by
\begin{equation}
p_{\phi}=\epsilon \frac {1}{2} \dot{\phi}^2- V(\phi)\,, \label{fi-pressure}
\end{equation}
\begin{equation}
\rho_{\phi}= \epsilon \frac {1}{2} \dot{\phi}^2+ V(\phi)\,. \label{fi-density}
\end{equation}
These two expressions  define an effective equation of state
$w_\phi=\displaystyle \frac{p_{\phi}}{ \rho_{\phi}}$, which drives
the  behavior of the model.  The field equations are
\begin{equation}
2\frac{\ddot{a}}{a}+ H^2 +\epsilon \frac {1}{2} \dot{\phi}^2 - V(\phi)=0,
\end{equation}
\begin{equation}
\frac{\ddot{a}}{a} + 3 H \dot{\phi}+\epsilon V^{\prime }(\phi)=0,
\end{equation}
\begin{equation}
 3H^2 = \rho_{\phi}+\rho_{m},
\end{equation}
where prime denotes derivative with respect to $\phi$, while dot denotes derivative with respect to time.

The Noether theorem states that, if there exists a vector field
$X$, for which the Lie derivative of a given Lagrangian $L$
vanishes i.e. $L_{X}{\cal L}=0$, the Lagrangian admits a Noether
symmetry and thus yields a conserved current \cite{cimento}. In
the Lagrangian under consideration, the configuration space is
${\cal M}=\{a,\phi \}$ and the corresponding tangent space is
$T{\cal M}=\{a,\phi ,\dot{a},\dot{\phi}\}$. Hence the
infinitesimal generator of the Noether symmetry is
\begin{equation}
X=\alpha \frac{\partial }{\partial {a}}+\beta \frac{\partial}{\partial {\phi }}+\dot{\alpha}\frac{\partial }{\partial {\dot{a}}}+\dot{\beta}\frac{\partial }{\partial {\dot{\phi}}},
\end{equation}
where $\alpha $ and $\beta $ are both  functions of $a$ and $\phi$
and
\begin{eqnarray}
&& \dot{\alpha} \equiv \frac{\partial\alpha }{\partial a}\dot{a} + \frac{\partial\alpha }{\partial \phi }\dot{\phi}\\
&&\dot{\beta}\equiv \frac{\partial \beta }{\partial a}\dot{a}+\frac{\partial \beta }{\partial \phi}\dot{\phi}.
\end{eqnarray}
The Cartan one--form is

\begin{equation}
\theta _{\cal L}=\frac{\partial {\cal L}}{\partial {\dot{a}}}da +
\frac{\partial {\cal L}}{\partial {\dot{\phi}}}d\phi .
\end{equation}
The constant  of motion  $Q=i_{X}\theta _{\cal L}$ is given by
\begin{equation}
Q=\alpha (a,\phi )\frac{\partial {\cal L}}{\partial
{\dot{a}}}+\beta (a,\phi)\frac{\partial {\cal L}}{\partial
{\dot{\phi}}}.
\end{equation}
If we demand the existence of a Noether symmetry, $L_{X}{\cal
L}=0$, we get the following equations,
 \begin{eqnarray}\label{eqnoether}
&&\alpha + 2 a \frac{\partial {\alpha }}{\partial {a}}=0\\
&& 6\frac{\partial {\alpha }}{\partial {\phi}} - \epsilon a^2\frac{\partial {\beta }}{\partial {a}}=0\\
&& 3 \alpha + 2 \epsilon a \frac{\partial {\beta }}{\partial {\phi}}=0\\
 && 3 V(\phi) \alpha + a V^{' }(\phi) a \beta =0\label{eqnoetherlast}
\end{eqnarray}
We have now to look for conditions on the integrability of this
set of equations, limiting ourselves to the \textit{phantom} case
(1.e., $\epsilon=-1$), since the standard case has been already
investigated in \cite{rubscud,marek1}. It is possible to
assume that $\alpha $ and $\beta $ are separable (and non--null),
i.e.
\begin{equation}
\alpha (a,\phi )=A_{1}(a)B_{1}(\phi ),~~~\beta (a,\phi
 )=A_{2}(a)B_{2}(\phi
).  \label{eq:fact}
\end{equation}
This is not true in general but, in such a case, it is
straightforward to achieve a solution for the system
(\ref{eqnoether}-\ref{eqnoetherlast}). It is
\begin{eqnarray}\label{alfabeta}
&&\alpha=\frac{2 A\cos\left(\frac{1}{2}\sqrt{\frac{3}{2}}\phi\right)}{\sqrt{a}}\,,\\
&&\beta=\frac{-2 \sqrt{6}A \sin\left(\frac{1}{2}\sqrt{\frac{3}{2}}\phi\right)}{a \sqrt{a}}\,,\\
&&V(\phi)=V_0 \sin\left(\frac{1}{2}\sqrt{\frac{3}{2}}\phi\right)^2
\end{eqnarray}
which selects the Noether symmetry.

\section{Solutions from new coordinates and Lagrangian}
Once that $X$ is found, it is then possible to find a change of
variables $\{a,\phi\} \rightarrow\{u,v\}$, such that one of them
(say $u$, for example) is cyclic for the Lagrangian ${\cal L}$ in
Eq. (\ref{eq:lagrangian}), and the transformed Lagrangian produces
a reduced dynamical system which is generally solvable. Solving
the system of equations $i_{X} d u = 1$ and $i_X d v = 0$ (where
$i_X d u$ and $i_X d v$ are the contractions between the vector
field $X$ and the differential forms $d u$ and $d v$,
respectively), we obtain:
\begin{eqnarray}
&& a =(v + 9 A^2 u^2 )^{\frac{1}{3}}\label{newvariable1}\\
&& \phi = 2 \sqrt{2/3} \arccos{\frac{(3 A u )}{\sqrt{v + 9 A^2 u^2 }}}.\label{newvariable2}
\end{eqnarray}
Under this transformation, the Lagrangian takes the suitable form
\begin{equation}\label{newl}
{\cal L}= D + v V_0 + \frac{\dot{v}^2}{3 v} + 12 A^2 \dot{u}^2\,,
\end{equation}
where $u$ is the cyclic variable. The conserved current gives
\begin{equation}
Q=\frac{\partial {\cal L}}{\partial \dot{u}} = 24 A^2\dot{u}=B,
\end{equation}
which can be trivially integrated to obtain $u(t)=B t+ C$. We use
now the energy condition $E_{\cal L}=0$ to find $v$.  We obtain
the following differential equation
\begin{equation}\label{eqv}
\dot{v(t)}^2-3v(t)\left(D+V_0 v(t)-12 A^2\dot{u(t)}^2\right)=0\,.
\end{equation}
It is a  first order equation which for , i.e. can be factorized
into the form
\begin{equation}
\left(p-F_1\right)\left(p-F_2\right)=0,
\end{equation}
being $p=\dot{v(t)}$ and $F_i=F_i(t,v)$. We are then left with
solving two first-degree equations $p=F_i(t,v)$. Writing the
solutions to these first-degree equations as $G_i(t,v)=0$ the
general solution to Eq. (\ref{eqv}) is given by the product
$G_1(t,v)G_2(t,v)=0$. It turns out that\footnote{In the following
we can set $C=0$ in $u(t)=B t+ C$ without loosing generality.}:
\begin{equation}
v(t)=\frac{\exp{(-\sqrt{3 V_0} t )}}{16 V_0^2}\left(\exp{(\sqrt{3V_0} t )}+ 48 A^2 B^2 V_0 - 4 D V_0\right)^2.
\end{equation}
The substitution of the functions $a = a(u,v)$ and $\phi =
\phi(u,v)$ into  Eqs. (\ref{newvariable1},\ref{newvariable2})
yields
\begin{eqnarray}
a(t)&=& \left(9 \omega^2 t^2 + \frac{\left(  \exp{-\sqrt{3 V_0}t}\right) \left(\exp{\sqrt{ 3 V_0 } t}+ 48 \omega^2 V_0 - 4 D V_0 \right)^{2}}{16 V_0^2}\right)^{\frac{1}{3}}\label{eqa}\\
\phi(t)&=& 2 \sqrt{\frac{2}{3}} \arccos{\left(\frac{3 A B t}{\sqrt{9 \omega^2 t^2 + \frac{\left(  \exp{-\sqrt{3 V_0}t}\right) \left(\exp{\sqrt{ 3 V_0 } t}+ 48 \omega^2 V_0 - 4 D V_0 \right)^{2}}{16 V0^2}}}\right)},\label{eqphi}
\end{eqnarray}
where we have defined $\omega = A B$. Setting $a(0)=0$, we can
construct a relation among the integration constants $\omega $,
$D$ and $V_0$: actually it turns out that $D =\frac{1 + 48
\omega^2  V_0}{4 V0}$. To determine the integration constant
$\omega $, we set the present time $t_0 = 1$. This fixes the
time-scale according to the (formally unknown) age of the
Universe. That is to say that we are using the age of the
Universe, $t_0$, as a unit of time. We then set $a_0 = a(1) = 1$,
to obtain
\begin{equation}\label{omega}
\omega=\frac{\sqrt{8 + \frac{1}{V_0^2} - \cosh{\sqrt{3 V_0}}}}{V_0^2}.
\end{equation}
The two conditions specified above allow one to express all the
basic cosmological parameters in terms of  $V_0$, the constant
that determines the scale of the potential. It  is not directly
measurable. However we can also strongly constrain its range of
variability through its relation with the Hubble constant:
actually because of our choice of time unit, the expansion rate
$H(t)$ is dimensionless, so that our Hubble constant
$\widehat{H}_0=H(t_0)$ is clearly of order 1 and not (numerically)
the same as the $H_0$ that is usually measured in ${\rm km s^{-1}
Mpc^{-1}}$. Actually, we can consider the relation
\begin{equation} \label{ha}
h=9.9\frac{{\widehat{H}_0}}{\tau}\,,
\end{equation}
where, as usual,  $h=H_{0}/100$ and $\tau$ is the age of the
Universe in $Gy$. We see that $\widehat{H}_0$ fixes only the
product $h\tau$. In particular, following e.g. \cite{w}, we can
assume that $\tau = 13.73^{+ 0.16}_{-0.15}$, thus we get $h< 0.76$
for $\widehat{H}_0 \approx 1$. Since, according to our
parameterization, $\widehat{H}_0 = \frac{2 + 16 V_0^2 - 2
\cosh{\sqrt{3 V_0}} + \sqrt{3 V_0}\sinh{\sqrt{3 V_0}}}{24 V_0^2}$,
it is possible to \textit{constrain} the range of variability for
$V_0$, starting from $\widehat{H}_0 $. By means of these choices,
the exact solutions in Eqs. (\ref{eqa}) and (\ref{eqphi}) can be
used to construct all the relevant cosmological parameters. In
particular
\begin{eqnarray}\label{cosmologicalf}
\rho_{\phi}&=&-\frac{1}{2}\dot{\phi(t)}^2 + V(\phi)\,,\\
p_{\phi}&=&-\frac{1}{2}\dot{\phi(t)}^2 - V(\phi)\,,\\
w_{\phi}&=&\frac{-\frac{1}{2}\dot{\phi(t)}^2 + V(\phi)}{-\frac{1}{2}\dot{\phi(t)}^2 + V(\phi)}\,,\\
\Omega_{\phi}&=&\frac{\rho_{\phi}}{3H^2}\,.
\end{eqnarray}

\begin{figure}
\includegraphics{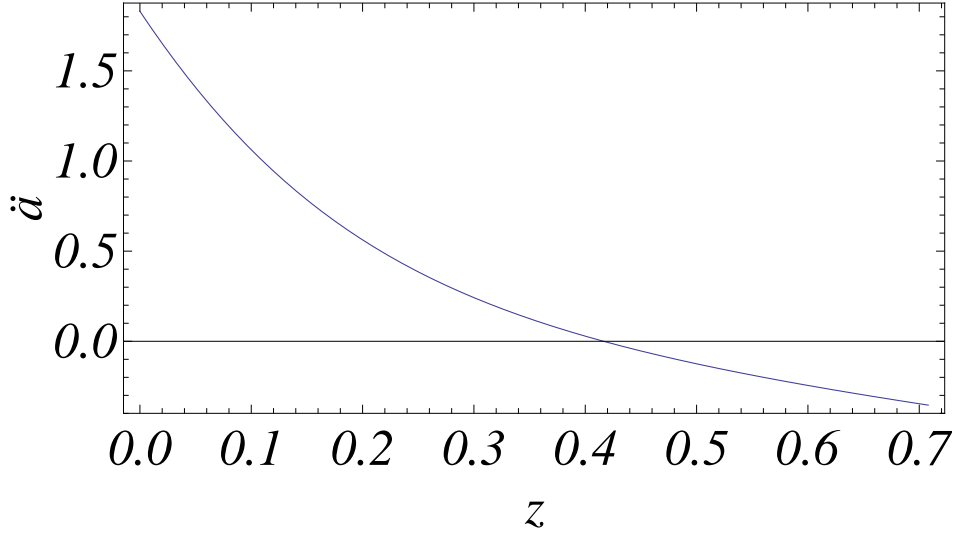}
\caption{ Redshift dependence of the acceleration $\ddot{a}(t)$:
the model allows an accelerated phase of expansion, as indicated
by the observations.} \label{fig:acc}
\end{figure}

\begin{figure}
        \includegraphics{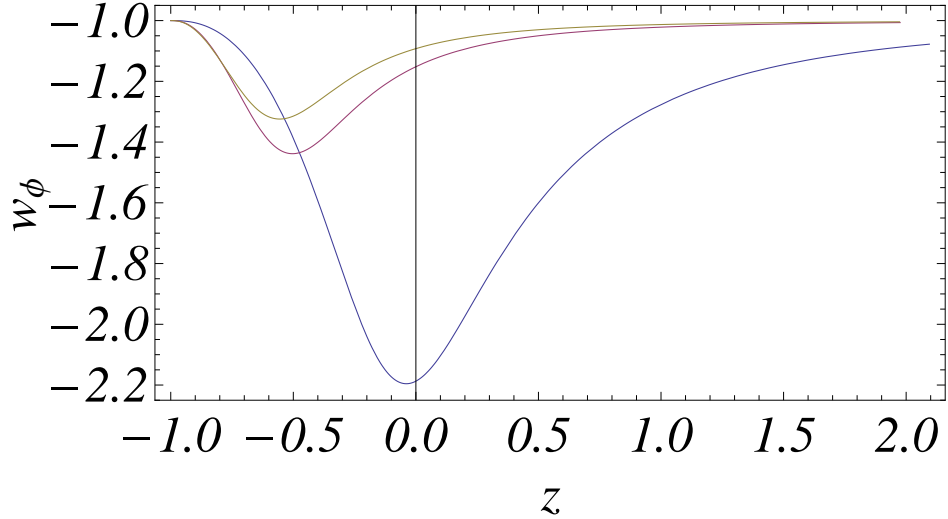}
        \caption{\small Redshift dependence of the equation of state parameter $w_\phi$
for the some values of $H_0$.} \label{fig:wz}
\end{figure}

\begin{figure}
\includegraphics{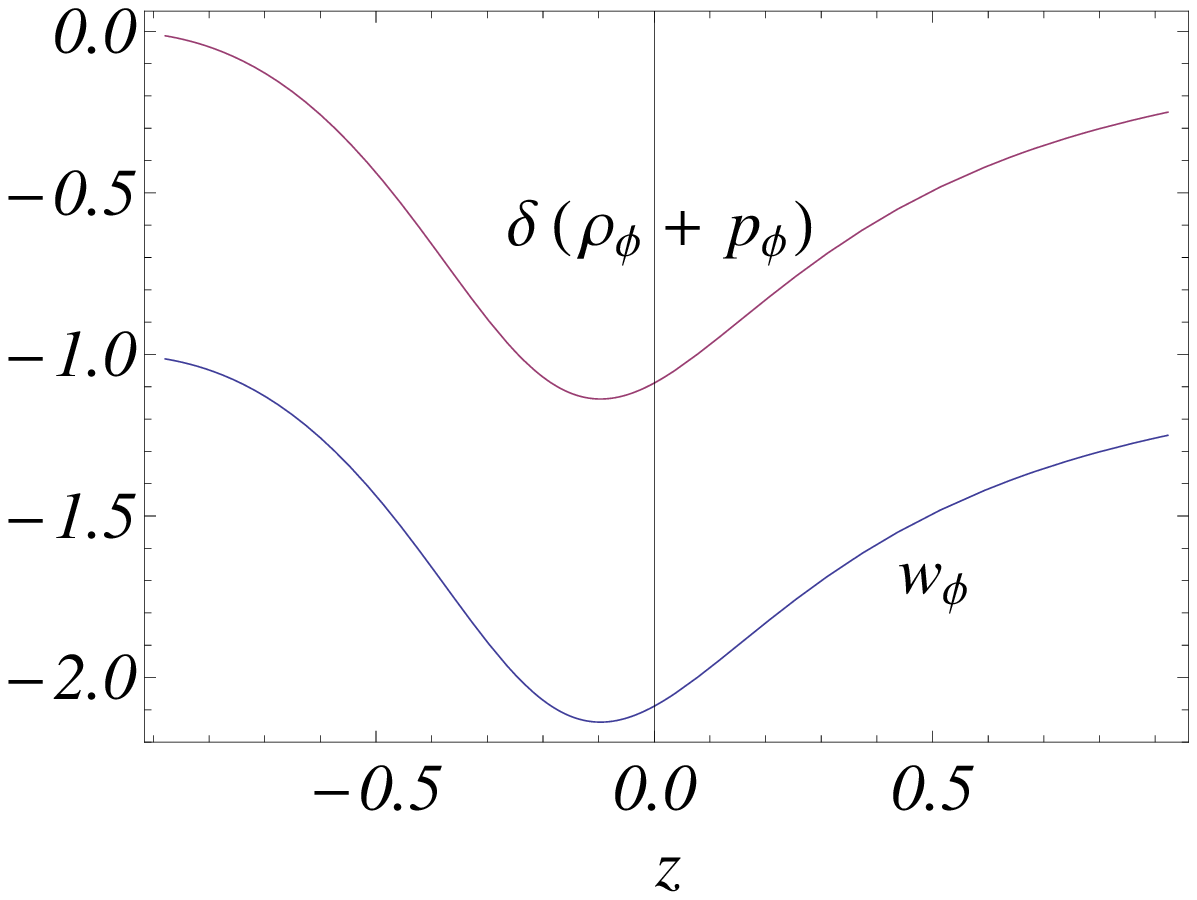}
\caption{\small Redshift dependence of the equation of state
$w_\phi$ and the function
$\delta(\rho_\phi+p_\phi)=\frac{\rho_\phi+p_\phi}{\rho_\phi}$,
which allows to compare the \textit{violation} of the energy
condition  with the \textit{super-quintessential} expansion. }
\label{fig:weakw}
\end{figure}
\begin{figure}
      \includegraphics{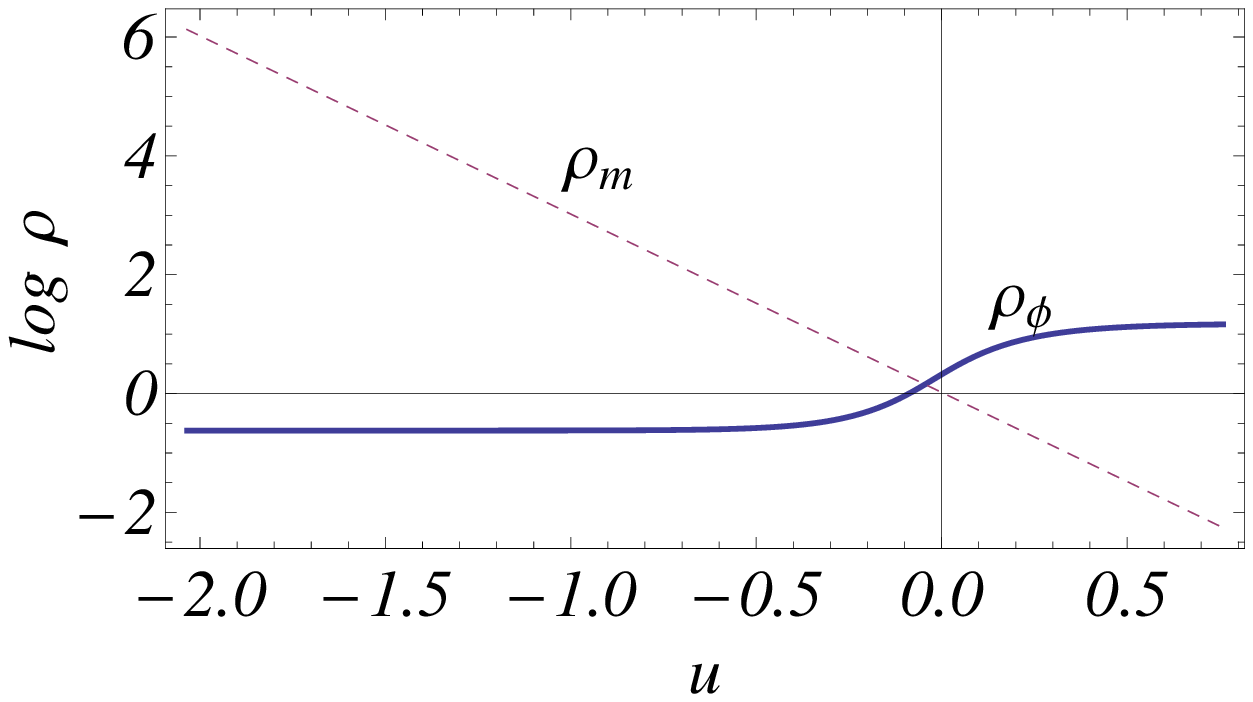}
        \caption{\small Plot of $\log_{10}{\rho_{\phi}}$ versus ${\log}_{10} a$ (thick line).
        The dashed lines indicate the log-log plot of $\log_{10}{\rho_{m}}\propto \log_{10}{a^{-3}} $. }
\label{fig:logplot}
\end{figure}

As it is shown in Fig. (\ref{fig:acc}), the model allows an
accelerated expansion as indicated from the observations, and
being a \textit{phantom} field exhibits a
\textit{super-quintessential} equations of state, with
$w_{\phi}<-1$ (see Fig. (\ref{fig:wz})), a violation of the weak
energy condition  (see Fig. (\ref{fig:weakw})). Finally, to
conclude this Section, we present the traditional plot
$\log{\rho_{\phi}}$ - $\log a$ compared with the matter density
(see Fig. (\ref{fig:logplot})). We see that $\rho_{\phi}$
undergoes a \textit{transition} from a subdominant phase, during
the matter dominated era, to a dominant phase. The nowadays
accelerated expansion of the Universe can be associated to such a
transition. Interestingly, both the subdominant and the dominant
phases are characterized by a constant density behavior.

\begin{figure}
        \includegraphics{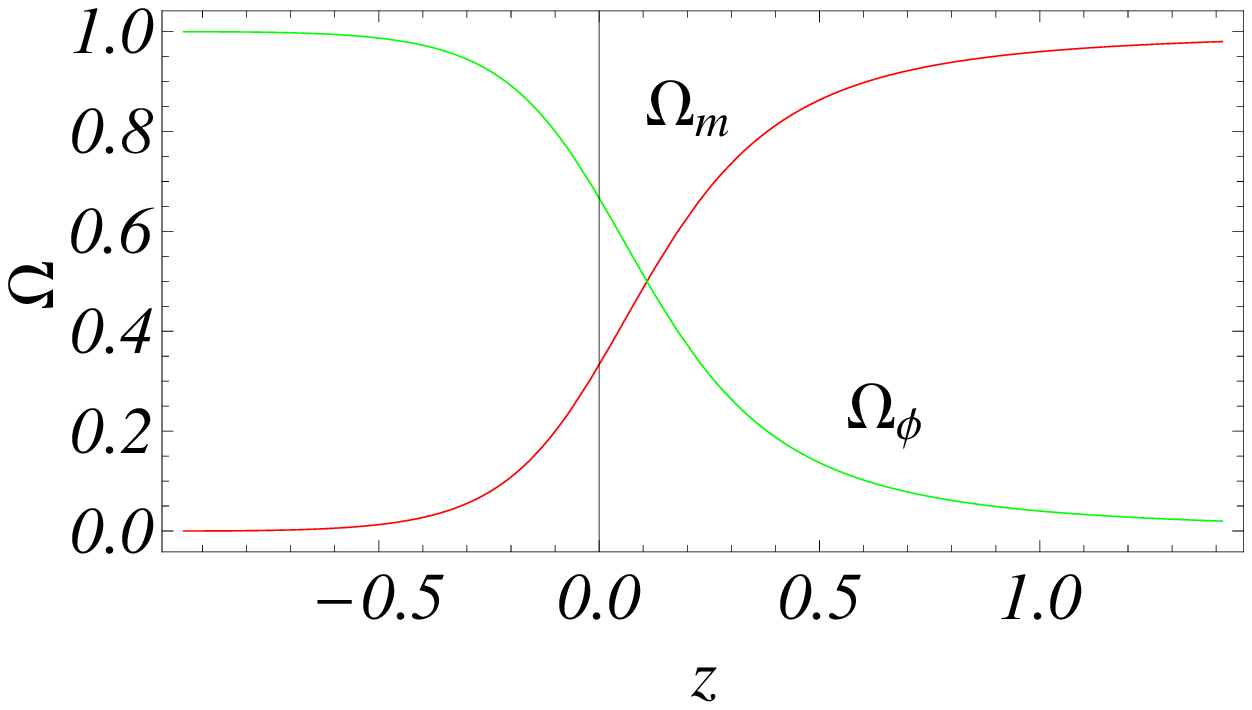}
        \caption{ Behavior of density parameters of (\textit{phantom}) quintessence and matter. The fixed point
regime, characterized by quintessence density parameter equal to
unity ($\Omega_\phi=1$), has not yet been reached today, which can
be considered a transition epoch.} \label{fig:omega}
\end{figure}
Fig.(\ref{fig:omega}) shows the density parameters of
(\textit{phantom}) quintessence and matter. The fixed point
regime, characterized by quintessence density parameter equal to
unity ($\Omega_\phi=1$), has not yet been  reached today. This
means that we are living in a transition epoch with
$\Omega_{\phi_0}\sim 0.7$.

\subsection{Including radiation}
A more realistic model can be considered by including also
radiation beside dust matter and scalar field. In this case, the
dynamical equations, as far as we know, do not have analytical
solutions and it is not possible to analytically reconstruct the
Noether symmetry. Due to these facts, we will relay on numerical
solutions.

\begin{figure}{
        \includegraphics{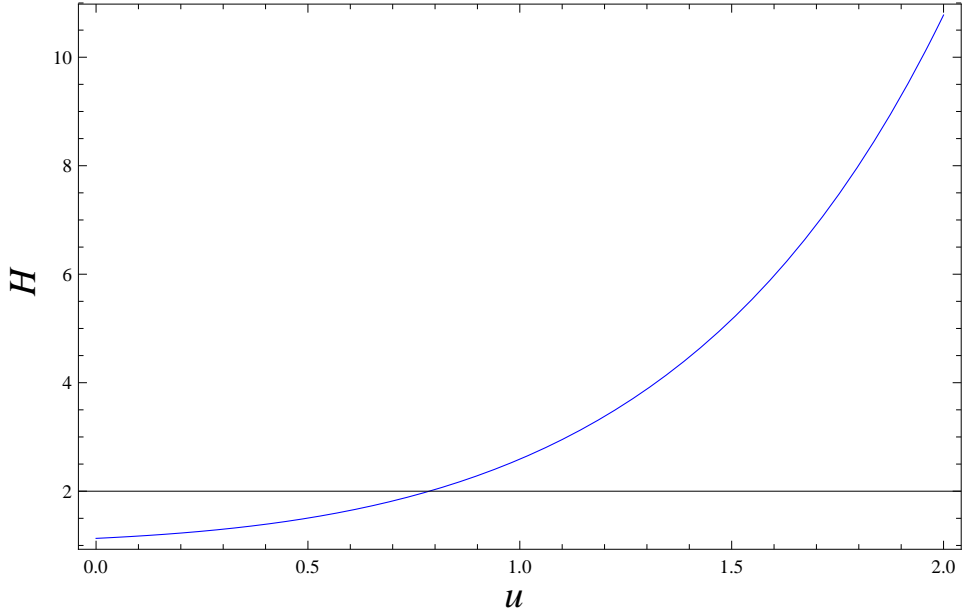}}
        \caption{The Hubble parameter as a function of $u$
        in the Universe filled in with matter, radiation and scalar field.}
        \label{hr}
\end{figure}

\begin{figure}{
        \includegraphics{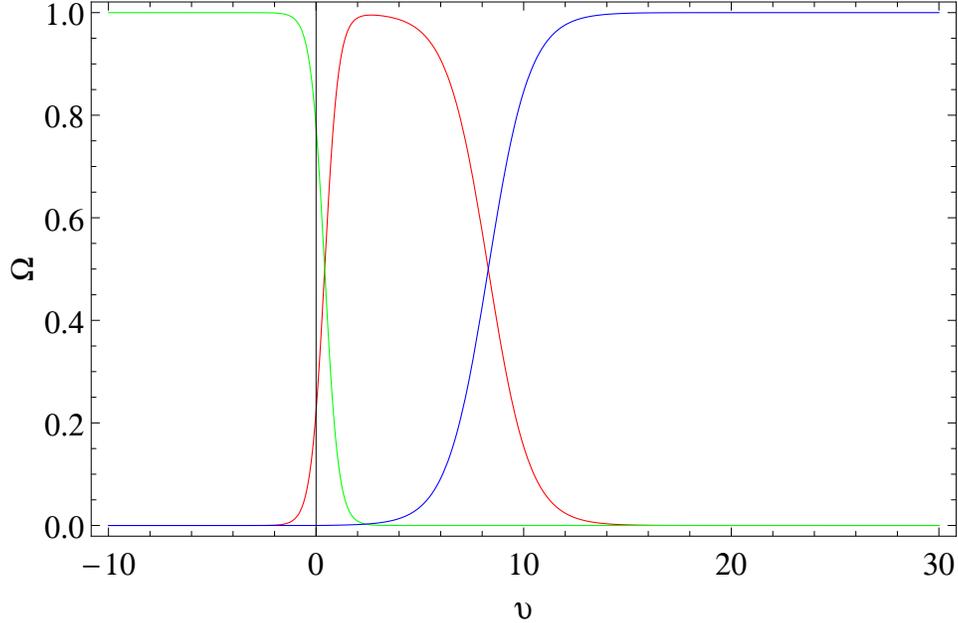}}
        \caption{Omega parameters as a function of $u$
        in the Universe filled in with matter, radiation and scalar field.
        $\Omega_{\varphi}$ is marked in green, $\Omega_{r}$ in red and $\Omega_{m}$ in blue.}
        \label{omegar}
\end{figure}

\begin{figure}
        \includegraphics{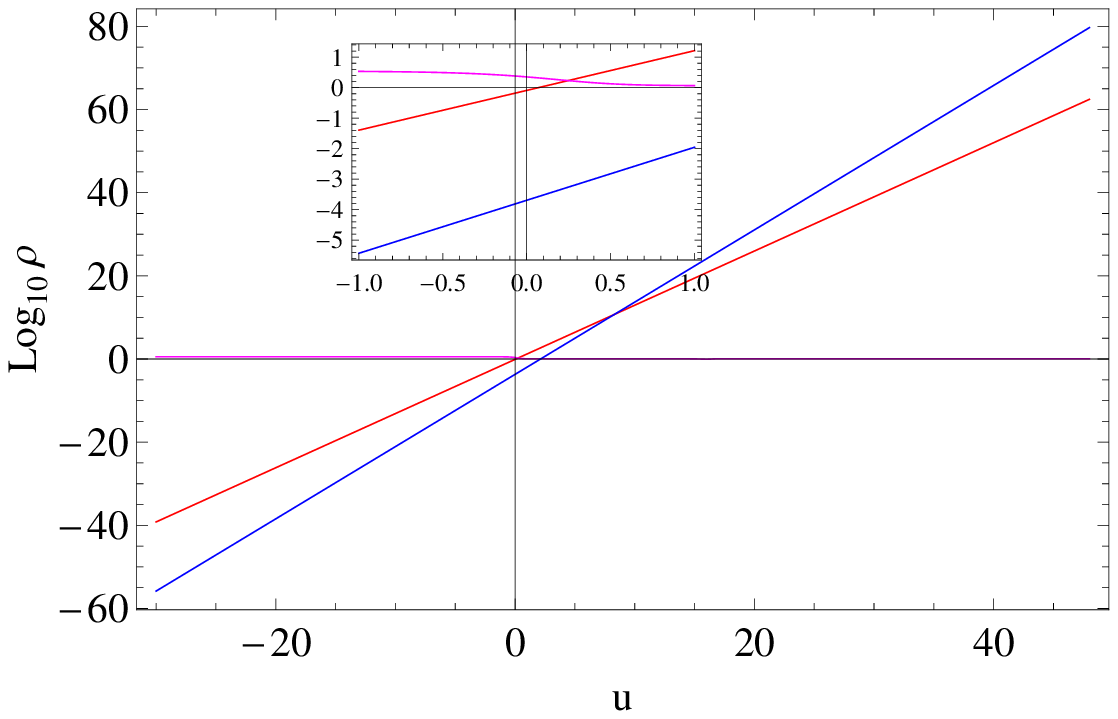}
        \caption{\small Plot of $\log_{10}{\rho_{\phi}}$ versus ${\log}_{10} a$ (thick line).
        The lines indicate the log-log plot of $\log_{10}{\rho_{m}}\propto \log_{10}{a^{-3}} $ and
the log-log plot of $\log_{10}(\rho_{r})\propto \log_{10}(a^{-4})$. }
\label{fig:logplotrad}
\end{figure}

Let us introduce the new independent variable
$u=\log(1+z)=-\log({a(t)/ a_{0}})$, where $a_{0}$ is the present
value of the scale factor. The Einstein scalar field equations can
be written in the form
\begin{equation}
 H^{2}= \frac{\rho_{m}+ \rho_{r}+ V}{ 3 + \frac{1}{2} {\phi}'^{2}}\,,\end{equation}
\begin{equation}
H^{2} \phi^{''}= \left[- \frac{1}{2}(\rho_{r}+\rho_{m})+
V\right]\phi'+ {\frac{d V}{d \phi}}\,\end{equation} where
$\rho_{r}\sim a^{-4}$ is the energy density of radiation. We
numerically solved this system of coupled equations, specifying
the initial conditions\footnote{The initial values are assumed so
that  $u=30$ gives the same value for $\phi(30)$, ${\phi(30)}'$
and $H(30)$. }. The results of numerical integration are shown in
Figs.\ref{hr}, \ref{omegar}, and \ref{fig:logplotrad}.

The presence of radiation is hardly changing the behavior of the
scalar field, its potential, the Hubble constant and the $w$
parameter of the dark energy equation of state. As expected, the
evolution of the $\Omega$ parameters is different. At the initial
time (fixed for numerical calculations at $u=30$), radiation
dominates the expansion rate of the Universe, with dark energy and
matter being sub--dominant. At a redshift of about 5000, the
energy density of matter and radiation become comparable and,
during a relatively short period, the Universe becomes matter
dominated. At a redshift of about $1$ dark energy starts to
dominate the expansion rate of the Universe. As result (see, Fig.
10), it follows that during the epoch of nucleosynthesis ($z\sim
10^{9}$) the energy density of the scalar field is much smaller
than the energy density of radiation. In particular, during such
an epoch, the kinetic terms in the energy density of scalar field
vanishes, and the potential terms is constant: in this case,  the
dark--energy term acts  as an effective cosmological constant
$\Lambda$, and it does not influence the process of primordial
nucleosynthesis.

\section{Observational data and predictions}
The above  \textit{phantom} scalar field model of quintessence,
provides an accelerated expansion of the Universe which could
agree, in principle, to the other cosmological behaviors. To test
the viability of the model, let us compare now its predictions
with some available observational dataset. We concentrate mainly
on different kinds of observational data: the publicly available
data on SNeIa, the parameters of large scale structure determined
starting from the 2-degree Field Galaxy Redshift Survey
(2dFGRS)and from the Wide part of the VIMOS-VLT Deep Survey
(VVDS).
\subsection{Constraints from  SNeIa observations}
The model can be constrained by SNeIa dataset presently available.
As a starting point,  let us take into account the sample of 182
SNIa compiled in \cite{Riess07}, which includes the 21 new SNeIa
recently discovered by the {\it Hubble Space Telescope (HST)}, and
combines previous SNeIa dataset, namely the Gold Sample compiled
in \cite{Riess04}, supplemented by the SNLS dataset \cite{SNLS}.

\begin{figure}
\includegraphics{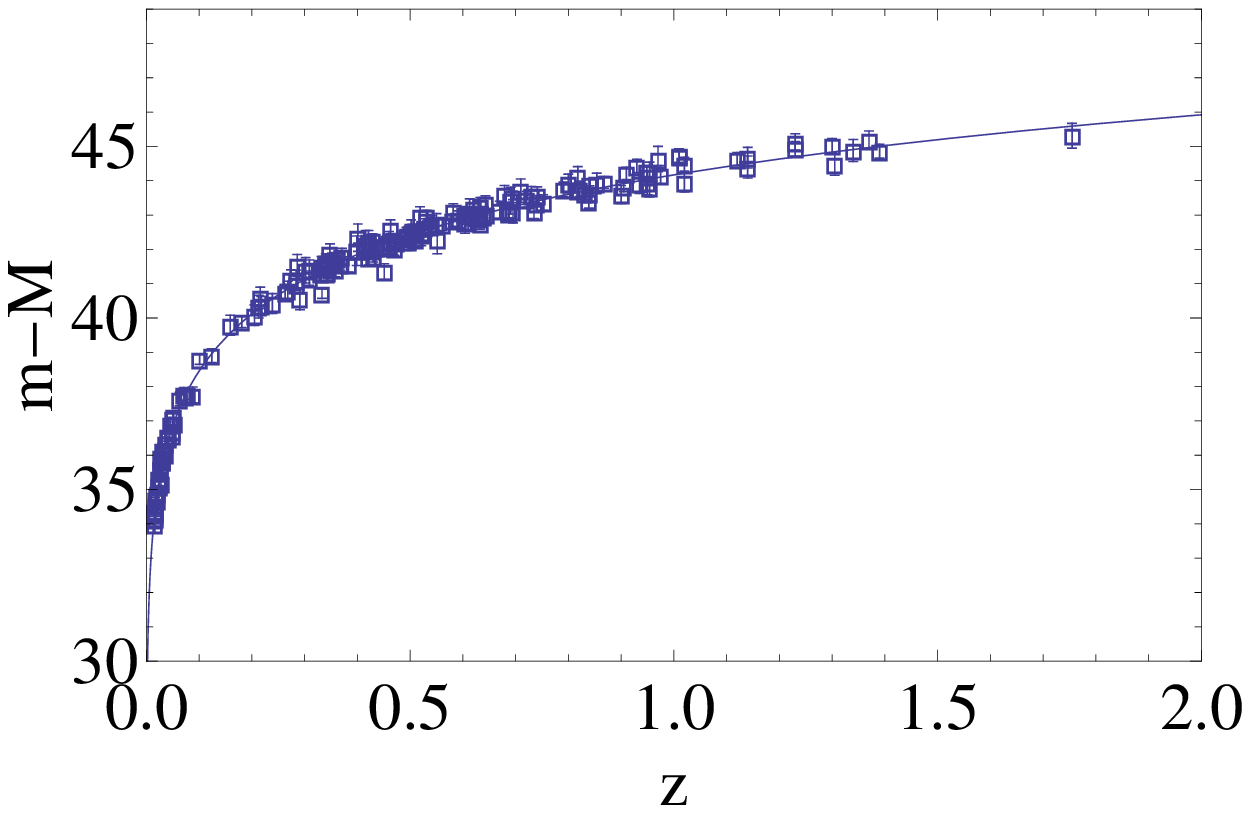}
        \caption{\small Observational data from the SNeIa sample fitted against the  model.
        The  solid curve is the best fit curve, and the best fits values are
 $V_0=14^{+2}_{-1}$, which corresponds
 to $\widehat{H}_0= 0.98^{+0.05}_{-0.04}$ and $\Omega_{\phi} = 0.68^{+0.06}_{-0.04}$. }
        \label{fig:bestfitsn}
\end{figure}

Following a standard procedure, we perform a
$\chi^2$ analysis comparing the redshift dependence of the
theoretical values to the observational estimates of the distance
modulus, $\mu=m-M$, which takes the form
\begin{equation}
m-M=5\log{D_{L}(z)}+  25.
\label{eq:modg}
\end{equation}
Moreover, the luminosity distance for a general flat and homogeneous cosmological model
can be expressed as an integral of the Hubble function as
\begin{eqnarray}\label{luminosity}
D_L (z) &=&
\frac{c}{H_0}(1+z)\int^{z}_{0}\frac{1}{E(\zeta)}d\zeta,
\end{eqnarray}
where $E(z)= \frac{H(z)}{H_0}$ is related to the Hubble function
expressed in terms of $z=a_0/a(t) - 1$. Let us note that the
luminosity distance also depends on the Hubble distance $c/H_0$
(which does not depend on the choice of the unit of time). Such
\textit{freedom} allows us to fit $h$ or the \textit{a priori}
unknown age of the Universe $\tau$ using the SNeIa dataset. We
find that $\chi_{red}^2=1.04$ for 182 data points, and the best
fit values are $V_0=14^{+3}_{-1}$, which corresponds to
$\widehat{H}_0= 0.98^{+0.05}_{-0.04}$ and $\Omega_{\phi } =
0.68^{+0.06}_{-0.04}$. We also get $h=0.72\pm 0.04$. In Fig
(\ref{fig:bestfitsn}), we compare the best-fit curve with the
observational dataset.
\subsection{Dimensionless coordinate distance test}
After having explored the Hubble diagram of SNeIa, that is the
plot of the distance modulus as a function of the redshift $z$, we
want here to follow a very similar, but more general approach,
considering as a cosmological observable the dimensionless
coordinate distance defined as\,:
\begin{equation}
y(z) = \int_{0}^{z}{\frac{1}{H(\zeta)} d\zeta} \ . \label{eq:
defy}
\end{equation}
The variable $y(z)$ does not depend explicitly on $h$ so that any
choice of $h$ does not alter the main result. Daly \& Djorgovki
\cite{DD04} have compiled a sample comprising data on $y(z)$ for
the 157 SNeIa in the Riess et al. \cite{Riess04} Gold dataset and
20 radiogalaxies from \cite{RGdata}, summarized in Tables\,1 and 2
of \cite{DD04}. In \cite{DD06}, they have added the latest SNeIa
data released from the SNLS collaboration \cite{SNLS} thus ending
up with a sample comprising 248 measurements of $y(z)$ that we use
here. As a preliminary step, Daly \& Djorgovski have fitted the
linear Hubble law to a large set of low redshift ($z < 0.1$) SNeIa
thus finding\,:
\begin{displaymath}
h = 0.664 \pm 0.008,
\end{displaymath}
which is consistent with our fitted value $h=0.72\pm 0.04$, and
with the value $H_0 = 72 \pm 8 \ {\rm km \ s^{-1} \ Mpc^{-1}}$
given by the HST Key project \cite{Freedman} based on the local
distance ladder and with the estimates coming from the time delay
in multiply imaged quasars \cite{H0lens} and the
Sunyaev\,-\,Zel'dovich effect in X\,-\,ray emitting clusters
\cite{H0SZ}. It is interesting to point out that the SHOES
Team recently completed an extensive new program with the Hubble
Space Telescope which stream-lined the old distance ladder and
observed Cepheids in the near-infrared  where they are less
sensitive to dust.  The result was to reduce the total uncertainty
in the Hubble constant by more than a factor of 2, now to just 4.8
percent uncertainty $(h=74.2 \pm 3.6)$ \cite{shoe}.

\begin{figure}{
        \includegraphics{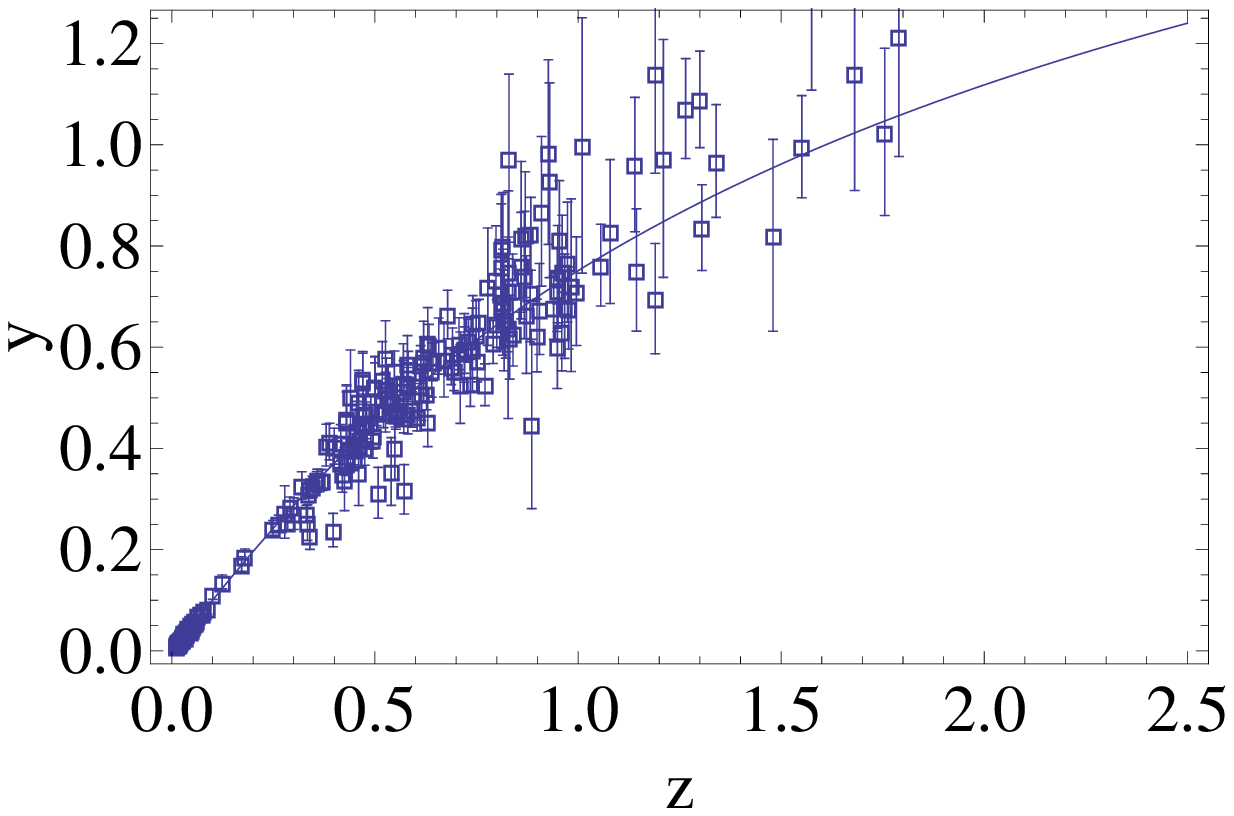}}
        \caption{\small Observational Daly \& Djorgovski database (\cite{DD06})
        fitted against the model. The solid curve is the best fit curve. }
        \label{fig:cord}
\end{figure}

\begin{figure}{
        \includegraphics{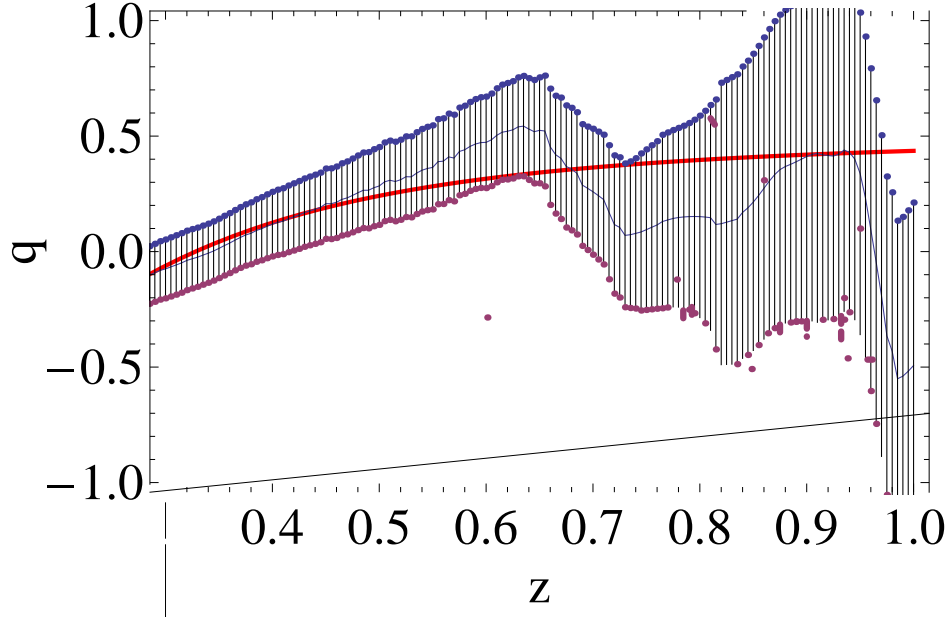}}
        \caption{ The allowed region for $q(z)$,
        obtained by Daly \& Djorgovski, from the full dataset (shadow area). An approximated
        polynomial corresponding to a $z$-window $\Delta z = 0.6$ is shown
        with the black--thin solid line.  With the black--thick dashed
        lines, it is shown the approximated polynomial fitted to the smoothed data  at
        $\pm 1\sigma$ range, and corresponding to a $z$-window $\Delta z = 0.4$.
        The red--solid line  shows the deceleration function, $q(z)$ for our
        model, corresponding to the the best fit values. }
        \label{dalyq}
\end{figure}

To determine the best fit parameters, we define the following
merit function\,:
\begin{equation}
\chi^2(V_0) = \frac{1}{N - 1} \sum_{i =
1}^{N}{\left [ \frac{y(z_i;  V_0) -
y_i}{\sigma_i} \right ]^2}\,. \label{eq: defchi}
\end{equation}
We obtain $\chi_{red}^2=1.1$ for 248 data points,  and the best
fit value is $V_0=14^{+ 3}_{-1}$, which corresponds to
$\widehat{H}_0= 0.96^{+0.1}_{-0.06}$ and $\Omega_{\phi } =
0.65^{+0.07}_{-0.04}$ In Fig (\ref{fig:cord}),  we compare the
best fit curve with the observational dataset.
Daly \& Djorgovski \cite{DD04} developed a numerical method for a
direct determination of the expansion and acceleration rates,
$H(z)$ and $q(z)$, from the data, using the dimensionless
coordinate distance $y(z)$, without making any assumptions about
the nature or evolution of the dark energy. They use the equation
\begin{equation}
-q(z) \equiv \ddot{a} a/\dot{a}^2 = 1~ +~ (1+z) ~(dy/dz)^{-1}
(d^2y/dz^2)\,,\label{dalyeq}
\end{equation}
valid for $k=0$. Eq.(\ref{dalyeq}) depends only upon the
Friedman-Robertson-Walker line element and the relation $(1+z) =
a_0/a(t)$. Thus, this expression for $q(z)$ is valid for any
homogeneous and isotropic Universe in which $(1+z) = a_0/a(t)$,
and it is therefore quite general and can be compared with any
model to account for the acceleration of the Universe. This
approach has the advantage to be model independent, but it
introduces larger errors in the estimation of $q(z)$, since the
numerical derivation is very sensitive to the size and quality of
 data. An additional problem is posed by the sparse and not
complete coverage of the $z$-range of interest. Measurement errors
are propagated in the standard way leading to estimated
uncertainties of the fitted values. In Fig. (\ref{dalyq}), we
compare the $q(z)$ obtained by Daly \& Djorgovski from their full
dataset with our {\it best fit} model.

\subsection{Growth of density perturbations and observational constraints from galaxies redshift surveys}
A relevant consequence of the presence of a dominant form of dark
energy in the Universe, in addition to its primary effect on the
expansion rate, is to modify the gravitational assembly of matter
from which the observed large-scale structure originated. In
linear perturbation theory, it is possible to describe the growth
of a generic small amplitude density fluctuation $\delta_M \equiv
\delta \rho_m /\rho_m$ through a second-order differential
equation \cite{peeb80,ma+al99}:
\begin{equation}
\label{grow1} \ddot{\delta}_m+2H(t)\dot{\delta}_m-4\pi{\rm G }\rho_m
\delta_m=0.
\end{equation}
In Eq.~(\ref{grow1}), the dark energy enters through its influence
on the expansion rate $H(t)$. We shall consider Eq.~(\ref{grow1})
only in the matter dominated era, when the contribution of
radiation is really negligible. In our model, Eq.(\ref{grow1})
assumes the form
\begin{eqnarray}
\label{grow2}
&&\ddot{\delta}_m + 2 \dot{\delta}_m \frac{2}{3}\,\left( \frac{-2t \left(-1 - 8 V_0^2 + \cosh{\sqrt{3V_0}}\right)+\sqrt{3 V_0}\sinh{\sqrt{3V_0}t}}{ -1 + t^2 + 8 t^2 V_0^2 - t^2 \cosh{\sqrt{3 V_0}} + \cosh{\sqrt{3V_0}t}} \right)+\nonumber\\
\\
&& -\delta_m \left(\frac{2 + V_0 (3 + 16 V_0) - 2 \cosh{\sqrt{3
V_0}}}{ -1 + t^2 + 8 t^2 V_0^2 - t^2 \cosh{\sqrt{3 V_0}} +
\cosh{\sqrt{3V_0}t}}\right)=0\,.
\end{eqnarray}
Eq.~(\ref{grow2}) does not admit exact analytic solutions.
However, since with our choice of normalization the whole history
of the Universe is confined in the range $t\in[0,1]$, and since we
choose $\omega \leq 1$, we can expand  the trigonometric functions
appearing in  Eq.~(\ref{grow2}) in series around $t=0$, obtaining
an integrable differential equation, which is a \textit{Fuchsian}
differential equation which admits hypergeometric solutions. For
the growing mode, we get
\begin{eqnarray}\label{growmode2}
&&\delta_+\propto t^{2/3}\ _2F_1 \left[
-\frac{1}{3},\frac{7}{6},\frac{11}{6}; -\frac{9 t^2 V_0^2}{
   24 + 36 V_0 + 192 V_0^2 - 24 \cosh{\sqrt{3V_0}}}\right]\,.
\end{eqnarray}
We use  such an exact solution to study the behavior of the
solution for $t\simeq 0$, and, mainly, to set the initial
conditions at $t=0$ and numerically integrate  Eq. (\ref{grow2})
in the whole range $[0,1]$. From its solutions, we can define a
linear growth rate $f$ that measures how rapidly structures are
being assembled in the Universe as a function of cosmic time, or,
equivalently, of the redshift:
\begin{equation}
\label{grow} f \equiv \frac{d \ln \delta_+ }{d \ln a}\,,
\end{equation}
where $a$ is the scale factor.

\begin{figure}{
        \includegraphics{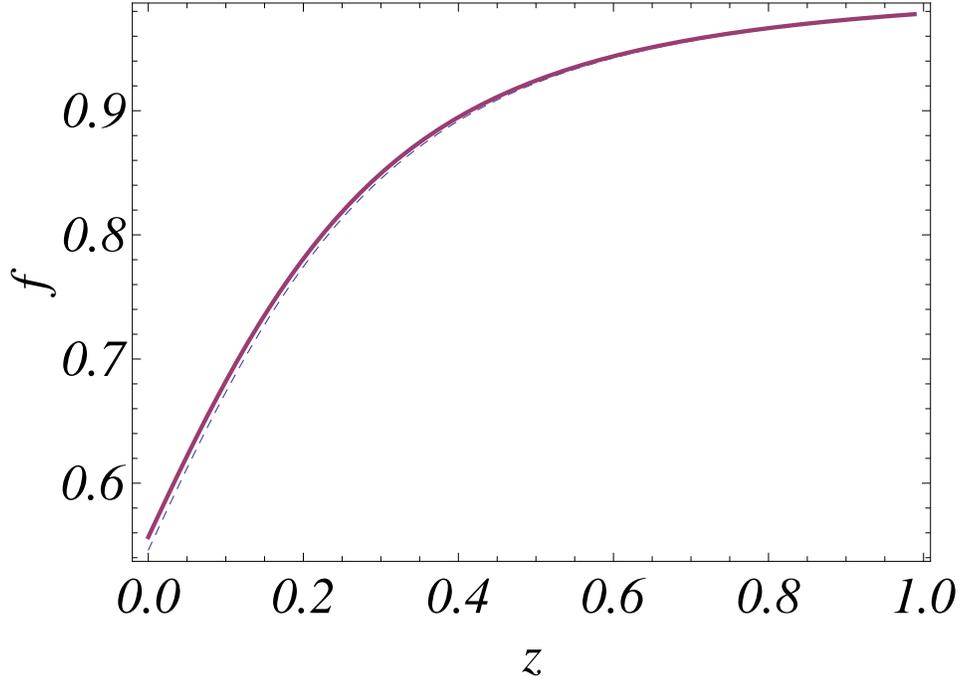}}
        \caption{\small The growth index $f$ in our cosmological model
        (the solid line), compared with its functional dependence from $\Omega_m^{\gamma}(z)$,
        with $\gamma=0.55$ (dashed line).}
        \label{growindex}
\end{figure}

\begin{figure}{
        \includegraphics{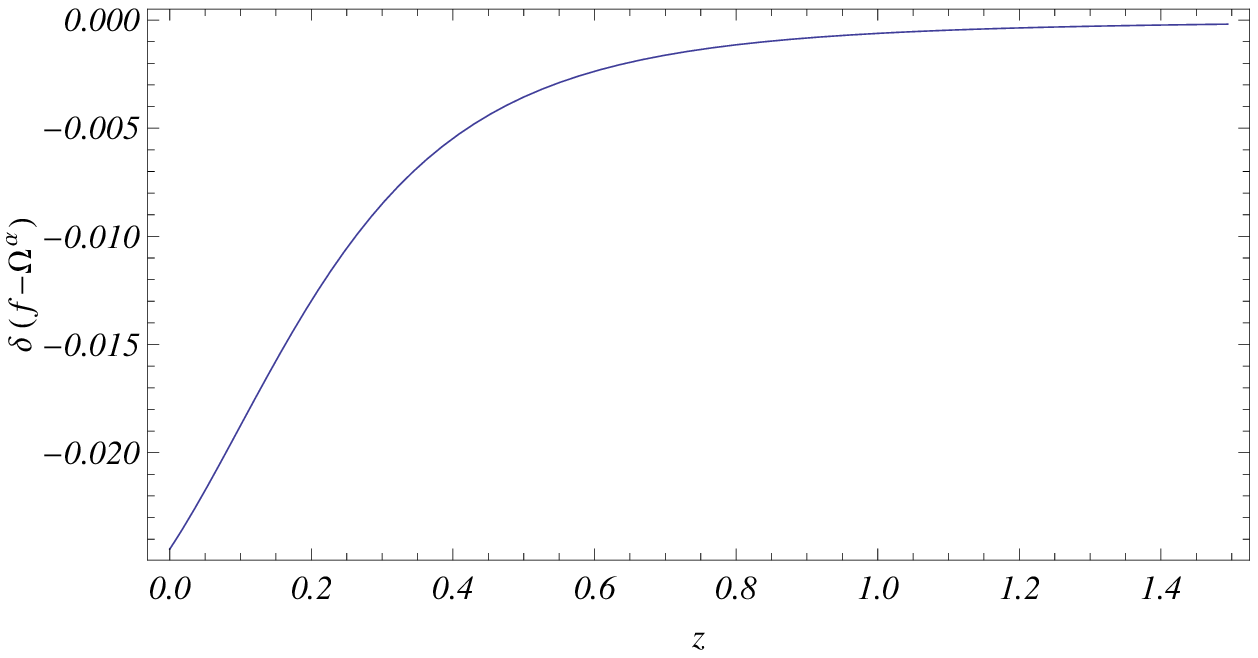}}
        \caption{\small The relative error between the growth index $f$ in our cosmological model
        $\Omega_m^{\gamma}(z)$, with $\gamma=0.55$: it turns out that such a
relation describes quite well the linear growth rate.}
        \label{grovariationomega}
\end{figure}

The growth index $f(z)$ essentially depends on the value of the
mass density parameter at the given epoch, $\Omega_m$. For the
cosmological-constant model the dependence is $f\propto
\Omega_m^{0.55}$. However, this is not valid if the observed
acceleration originates from a modification of the equations of
General Relativity; for example, in the Dvali-Gabadadze-Porrati
(DGP) braneworld theory \cite{DGP,DGP2,DGP3}, an extra-dimensional
modification of gravity gives $f(z)\propto \Omega_m^{0.68}(z)$. In
general, a fitting form $f(z)f(z)\propto \Omega_m^{\gamma}(z)$ has
been shown to be an accurate description for a wide range of
models (for which $\Omega_m^{\gamma}(z)$ itself, not only
$\gamma$, depends on the model). Thus, models with the same
expansion history $H(z)$ but a different gravity theory, will have
a different growth rate evolution $f(z)$ and index $\gamma$. A
discrepancy between the measured value of the growth rate and that
computed independently (assuming General Relativity) from the
$H(z)$ yielded by SNeIa would point out modifications of gravity.
In Figs. \ref{growindex}, \ref{grovariationomega}, we show that,
for the parameters of our model, the relation $f(z)\propto
\Omega_m^{0.68}(z)$ works quite well. Some observational
techniques have been suggested to measure $f(z)$ at different
redshifts.  Redshift-space distortions, that is, the imprint of
large-scale peculiar velocities on observed galaxy maps, have not
yet been considered in this context. Gravity driven coherent
motions are in fact a direct consequence of the growth of
structure. The anisotropy they induce in the observed galaxy
clustering, when redshifts are used as a measure of galaxy
distances, can be quantified by means of the redshift-space
two-point correlation function $\xi(r_p,pi)$,where $r_p$ and $\pi$
are respectively the transverse and line--of--sight components of
galaxy separations. The anisotropy of $\xi(r_p,pi)$ has a
characteristic shape at large $r_p$ that depends on the parameter
$\beta = \frac{f}{ b_L}$. In practice, we observe a compression
that is proportional to the growth rate, weighted by the factor
$b_L$, the linear bias parameter of the specific class of galaxies
being analyzed. The parameter $b_L$ measures how closely galaxies
trace the mass density field, and is quantified by the ratio of
the root-mean-square fluctuations in the galaxy and mass
distributions on linear scales. A value of $\beta 0.49 \pm 0.09$
has been measured at $z =0.15$ using the 2dF Galaxy Redshift
Survey (2dFGRS) sample of 220,000 galaxies with bias
\cite{ver+al01,la+al02}. From the observationally determined
$\beta$ and $b$, it is now straightforward to get the value of the
growth index at $z=0.15$ corresponding to the effective depth of
the survey. Verde \& al.~(2001) used the bi-spectrum of 2dFGRS
galaxies, and Lahav \& al. ~(2002) combined the 2dFGRS data with
CMB data, and they obtained
\begin{eqnarray}\label{bias}
 b_{L_{verde}}&=&1.04\pm 0.11\,,\\
 b_{L_{lahav}}&=&1.19\pm 0.09\,.
\end{eqnarray}

Using these two values for $b$  the value of the growth index
$f$ at $z=0.15$ is
\begin{eqnarray}\label{peculiar}
 f_1&=&0.51\pm 0.1\,,\\
 f_2&=&0.58 \pm 0.11\,.
\end{eqnarray}
More recently,  Guzzo et al.  reported a measurement of $\beta=
0.70 \pm 0.26$ at a redshift of $0.77$, using new spectroscopic
data from the Wide part of the VIMOS-VLT Deep Survey (VVDS)
\cite{guzzo08}. Using a new survey of more than 10,000 faint
galaxies, they also measured the anisotropy parameter $b_L = 0.70
\pm 0.26$, which corresponds to a growth rate of structure at that
time of $f = 0.91 \pm 0.36$. This is consistent with our
cosmological model, which gives $f(0.77)=0.97\pm 0.12$and
$f(0.15)=0.68\pm 0.2$. However it is also consistent  with
standard $\Lambda$CDM with low matter density and flat geometry,
although the error bars are still too large to distinguish among
alternative origins for the accelerated expansion. This could be
achieved with a further factor-of-ten increase in the sampled
volume at similar redshift.

\section{Conclusions}
We have  investigated the possibility that phantom field dynamics
could be derived by the Noether Symmetry Approach. The method
allows to fix the self--interacting potential of the phantom field
and then to solve exactly the field equations, at least in the
case of dark energy and matter dominated Universe. The main
cosmological parameters can be directly derived starting from the
general solution. We also worked out a comparison between the
theoretical predictions  and observational dataset, as the
publicly available data on SNeIa and radiogalaxies, the parameters
of large scale structure determined from the 2-degree Field Galaxy
Redshift Survey (2dFGRS)and from the Wide part of the VIMOS-VLT
Deep Survey (VVDS). It turns out that the model is quite well
compatible with the presently available observational data.

Furthermore, we extended the approach to the case  including
radiation. It can be shown that radiation is hardly changing the
behavior of the scalar field, its potential, the Hubble constant
and the $w$ parameter of the dark energy equation of state. As
expected, the evolution of the $\Omega$ parameters is different.
At the initial epochs, radiation dominates the expansion rate of
the Universe, with dark energy and matter being sub-dominant. At a
redshift of about 5000, the energy density of matter and radiation
become almost equivalent and, for a relatively short period, the
Universe becomes matter dominated. At a redshift of about $1$,
dark energy starts to dominate the expansion rate of the Universe.
It turned out that during the epoch of nucleosynthesis ($z\sim
10^{9}$) the energy density of the scalar field is much smaller
than the energy density of radiation, and during such an epoch the
kinetic terms in the scalar--field energy density vanishes, and
the potential terms is constant. This meas that the dark-energy
term acts  as an effective cosmological constant $\Lambda$, and it
does not influence the process of primordial nucleosynthesis.

As concluding remark, it is interesting to see that the presence
of the Noether symmetry could constitute a physical criterion to
fix the phantom potential. Such an approach revealed extremely
useful also for other classes of models (see, e.g.
\cite{defelice,leandros}).

\end{document}